\documentclass[letter]{aa}  

\usepackage{graphicx}
\usepackage{txfonts}

\makeatletter
\renewcommand*\aa@pageof{, page \thepage{} of \pageref*{LastPage}}
\makeatother

\usepackage[colorlinks=true, allcolors=blue]{hyperref}

\begin{document}

   \title{Intrinsic line profiles for X-ray fluorescent lines in SKIRT}

   \titlerunning{Intrinsic line profiles in SKIRT}

   \authorrunning{B.\ Vander Meulen et al.}

   \author{Bert Vander Meulen
          \inst{1}
          \and
          Peter Camps
          \inst{1}
          \and
          Masahiro Tsujimoto
          \inst{2}
          \and
          Keiichi Wada
          \inst{3,4,5}
          }
   \institute{Sterrenkundig Observatorium, Universiteit Gent, Krijgslaan 281 S9, 9000 Gent, Belgium\\
              \email{bert.vandermeulen@ugent.be}
              \and
              Institute of Space and Astronautical Science, Japan Aerospace Exploration Agency, Kanagawa 252-5210, Japan
              \and
              Graduate School of Science and Engineering, Kagoshima University, Kagoshima 890-0065, Japan
              \and
              Research Center for Space and Cosmic Evolution, Ehime University, Matsuyama 790-8577, Japan
              \and
              Faculty of Science, Hokkaido University, Sapporo 060-0810, Japan
        }
   \date{Received July 3, 2024; accepted August 4, 2024}


  \abstract
   {X-ray microcalorimeter instruments are expected to spectrally resolve the intrinsic line shapes of the strongest fluorescent lines. X-ray models should therefore incorporate these intrinsic line profiles to obtain meaningful constraints from observational data.}
   {We included the intrinsic line profiles of the strongest fluorescent lines in the X-ray radiative transfer code SKIRT to model the cold-gas structure and kinematics based on high-resolution line observations from XRISM/Resolve and \emph{Athena}/X-IFU.}
   {The intrinsic line profiles of the $\mathrm{K}\alpha$ and $\mathrm{K}\beta$ lines of Cr, Mn, Fe, Co, Ni, and Cu were implemented based on a multi-Lorentzian parameterisation. Line energies are sampled from these Lorentzian components during the radiative transfer routine.}
   {In the optically thin regime, the SKIRT results match the intrinsic line profiles as measured in the laboratory. With a more complex 3D model that also includes kinematics, we find that the intrinsic line profiles are broadened and shifted to an extent that will be detectable with XRISM/Resolve; this model also demonstrates the importance of the intrinsic line shapes for constraining kinematics. We find that observed line profiles directly trace the cold-gas kinematics, without any additional radiative transfer effects.
   }
   {With the advent of the first XRISM/Resolve data, this update to the X-ray radiative transfer framework of SKIRT is timely and provides a unique tool for constraining the velocity structure of cold gas from X-ray microcalorimeter spectra.}


   \maketitle
%
\section{Introduction}
\label{sect:intro}
Fluorescent lines are the most prominent tracers of X-ray reprocessing by cold atomic gas and can be observed in the X-ray spectra of active galactic nuclei (AGNs), X-ray binaries, and more \citep{george91, matt91, nandra94, torrejon10}. These lines encode physical information about the reprocessing medium and are commonly used to constrain the structure of cold gas surrounding X-ray sources \citep{murphy09, ikeda09, brightman11, odaka11, liu14, paltani17, balokovic18, buchner19, vandermeulen23}.

Fluorescent lines are produced when an inner-shell electron is ejected from a gas atom by the absorption of an X-ray photon. A higher-shell electron can fall into the inner-shell vacancy, emitting the energy difference as a fluorescent line. The strongest fluorescent lines correspond to the following K-shell transitions, 
\begin{align}
    &\mathrm{K}\alpha\textrm{: } \textrm{L}_3 \to \textrm{K} \; (\textrm{K}{\alpha_1}) \;\textrm{and L}_2 \to \textrm{K} \;(\textrm{K}{\alpha_2}), \notag\\
    &\mathrm{K}\beta \textrm{: } \textrm{M}_3 \to \textrm{K} \;(\textrm{K}{\beta_1})   \;\textrm{and M}_2 \to \textrm{K} \;(\textrm{K}{\beta_3}) ,\notag
\end{align}
and are most prominently observed for abundant heavy metals, such as Fe and Ni, as the fluorescent yields increase along the atomic sequence. \citep[][]{perkins91}.

The natural line shape for these bound-bound transitions (in energy space) is a Lorentzian \citep{agarwal79}. The $\mathrm{K}\alpha$ doublet of most elements can be accurately represented with two Lorentzian sub-line components \citep{thomsen83}, whereas $\mathrm{K}\beta$ lines often resemble a single Lorentzian due to the proximity of the $\textrm{K}{\beta_1}$ and $\textrm{K}{\beta_3}$ sub-lines \citep[][]{perkins91}. However, for 3d transition elements (i.e.\ Sc to Zn), the K$\alpha$ and K$\beta$ sub-lines are significantly asymmetric due to multi-electron interactions inside the atom \citep{druyvesteyn28, tsutsumi68, doniach70, finster71, deutsch95}.

With the advent of X-ray microcalorimeter instruments such as XRISM/Resolve \citep[][]{tashiro20} and \emph{Athena}/X-IFU \citep[][]{barret18}, the intrinsic line shapes of the strongest fluorescent lines (Fe K$\alpha$, Fe K$\beta$, Ni K$\alpha$, and more) are expected to be spectrally resolved \citep{smith16, mizumoto22, cuccetti24, kilbourne24, vaccaro24}. Therefore, X-ray spectral models should incorporate these line profiles in order to derive meaningful (velocity) constraints from observational data.

The intrinsic line shapes of the K$\alpha$ and K$\beta$ lines of Cr, Mn, Fe, Co, Ni, and Cu were measured in the laboratory by \citet{holzer97} using a single-crystal diffractometer with a sub-part-per-million absolute-energy-scale accuracy. These data were presented alongside an analytical parameterisation for each of these lines, providing a convenient way to incorporate the line profiles into X-ray spectral models. Building on this parameterisation, the HEASARC calibration database\footnote{\url{https://heasarc.gsfc.nasa.gov/FTP/caldb/}} (CALDB) listed a minor correction for the Mn K$\alpha$ line based on onboard \emph{Hitomi}/SXS calibration measurements \citep{eckart16}.

The intrinsic line profiles of the Fe K$\alpha$ and Fe K$\beta$ lines were recently implemented into MYTORUS \citep{yaqoob24}. MYTORUS \citep{murphy09} is an X-ray spectral model for absorption and reflection by a toroidal re-processor of cold atomic gas, which can be used to model the `AGN torus'. \citet{yaqoob24} finds that the addition of these intrinsic line profiles is crucial for estimating the velocity broadening of the cold gas if the true velocity broadening is less than about $2000~\textrm{km s}^{-1}$.

This Letter reports on the implementation of the intrinsic line profiles of 12 fluorescent lines 
into the X-ray radiative transfer code SKIRT \citep{vandermeulen23}. SKIRT\footnote{\url{https://skirt.ugent.be}} is a state-of-the-art Monte Carlo radiative transfer framework that is developed and maintained at Ghent University \citep{baes03, baes11, camps15a, camps20}. Recently, the SKIRT code was extended into the X-ray range to study the complex distribution of cold gas and dust around AGNs \citep{vandermeulen23}. SKIRT is optimised to operate in arbitrary 3D geometries with support for 3D kinematics and covers the X-ray to millimetre wavelength range self-consistently \citep{camps20, gebek23, matsumoto23}. In the X-ray range, the SKIRT code models Compton scattering on free electrons, photo-absorption and (self-consistent) fluorescence by cold atomic gas, scattering on bound electrons, and extinction by dust. More details on the X-ray processes that are implemented in SKIRT are presented in \citet{vandermeulen23}.

With the advent of the first XRISM/Resolve data, this update to the X-ray framework of SKIRT is timely and will be crucial for constraining the 3D density and velocity structure of cold gas from microcalorimeter X-ray spectra. The outline for this Letter is as follows: In Sect.~\ref{sect:implementation} we present the implementation of the intrinsic line profiles. In Sect.~\ref{sect:verification} we compare the SKIRT simulation results to laboratory data. In Sect.~\ref{sect:demonstration} we demonstrate the capabilities of SKIRT to predict detailed line shapes from complex 3D models. We summarise our results and discuss their implications in Sect.~\ref{sect:summary}.

\section{Implementation}
\label{sect:implementation}
The intrinsic line profiles of the $\mathrm{K}\alpha$ and $\mathrm{K}\beta$ lines of Cr, Mn, Fe, Co, Ni, and Cu were measured by \citet{holzer97}, who provided an analytical parameterisation for each profile based on four to seven Lorentzian functions per line, which we adopted in SKIRT. The Lorentzian (i.e.\ Cauchy) probability density function is\footnote{Equation~1 of \citet{holzer97} is missing the factor $1/2$.}
\begin{equation}
    \label{eq:pdf}
    p(E) = I_i\,\,{\left[1 + \left(\frac{E-E_i}{W_i/2}\right)^2\right]}^{-1},
\end{equation}
with $E_i$ being the photon energy of the line centre, $W_i$ the full width at half maximum, and $I_i$ the amplitude of the line. The integral (norm) of this function is $I_\textrm{int} = \pi\, I_i \, W_i/2$. 

As the cumulative probability function of a Lorentzian can be analytically inverted, random photon energies can easily be generated from Eq.~\ref{eq:pdf} through inverse transform sampling as
\begin{equation}
    E = \frac{W_i}{2}\, \tan\left[ \pi \, \left(\chi - \frac{1}{2}\right)\right] +E_i,
    \label{eq:sample}
\end{equation}
with $\chi$ being a uniform deviate between 0 and 1.

The $E_i$, $W_i$, $I_i$, and $I_\textrm{int}$ parameters of all Lorentzian sub-line components are taken from \citet{holzer97} except for the Mn $\mathrm{K}\alpha$ sub-lines, which are taken from CALDB and include a minor correction to the Mn $\mathrm{K}\alpha$ line shape \citep{eckart16}. Furthermore, as the line integrals, $I_\textrm{int}$, of the Fe $\mathrm{K}\alpha_{22}$, Fe $\mathrm{K}\alpha_{23}$, and Co $\mathrm{K}\alpha_{14}$ sub-lines\footnote{This is also true for the Mn $\mathrm{K}\alpha_{15}$ and Mn $\mathrm{K}\alpha_{22}$ sub-line parameters as listed in \citet{holzer97}, but they are not used in this work.} are not consistent with their amplitudes, $I_i$, we recalculated the $I_\textrm{int}$ values based on the listed $I_i$ and $W_i$ parameters\footnote{These differences are likely not observable with XRISM/Resolve.} -- a choice that is consistent with the experimental Fe~$\mathrm{K}\alpha$ data shown in Fig.~1 of \citet{holzer97}. Finally, we normalised the $I_\textrm{int}$ and $I_i$ sub-line parameters so that each line (i.e.\ the sum of its sub-lines) has a total norm of 1.

While the Lorentzian sub-line components are merely part of an empirical parameterisation and do not represent any physical lines, they can be treated as individual lines in SKIRT, with an `artificial' fluorescent yield of $Y_{k,i} = I_\textrm{int} \times Y_k$, with $Y_k$ being the yield of the actual line \citep{perkins91}. In this way, most of the current implementation can be retained \citep[see][]{vandermeulen23}, with one key difference: when a new Lorentzian sub-line (with yield $Y_{k,i}$) is selected after a photo-absorption event, the corresponding line energy is no longer fixed (i.e.\ tabulated). Instead, a line energy must be sampled using Eq.~\ref{eq:sample}, employing the sub-line parameters $E_i$ and $W_i$ that correspond to that specific Lorentzian sub-line. The main advantage of this approach is that line photons are traced individually, instead of propagating the line profile in its entirety, so radiative transfer effects on the line profile can be modelled \citep{sukhorukov17}. As the brightest lines are more frequently sampled (more photons are emitted in these lines), we naturally obtain high-signal-to-noise line profiles for the most prominent fluorescent lines. We do not find a significant difference in the simulation runtime with or without intrinsic line profiles.

\begin{figure*}[!th]
    \centering
        \includegraphics[width=\hsize]{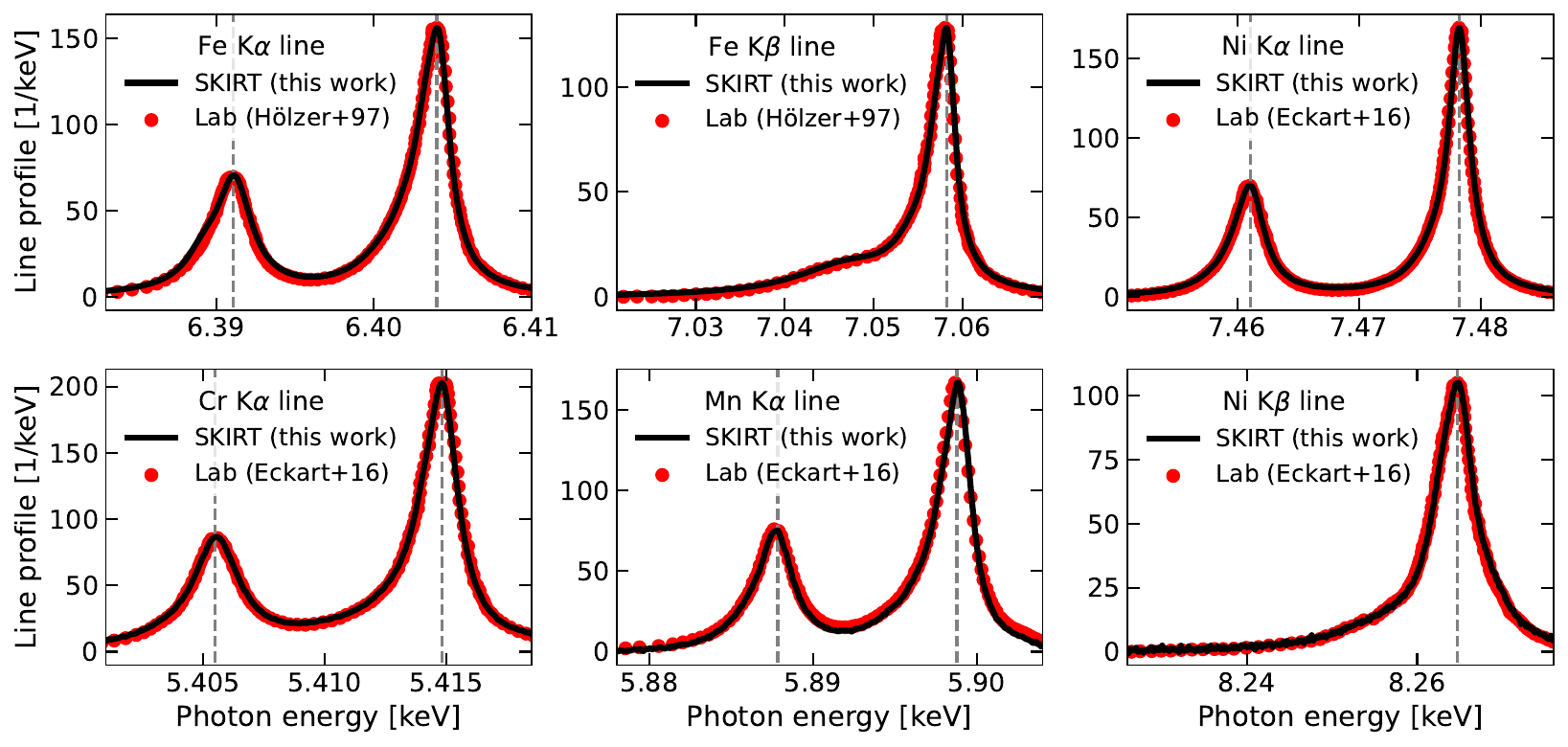}
    \caption{Continuum-subtracted line profiles for the strongest fluorescent lines as obtained with SKIRT in the optically thin regime (black) compared to the laboratory data of \citet{holzer97} (red), which have been corrected for instrumental broadening. We find an excellent agreement for all lines, verifying the SKIRT implementation and the data corrections that are described in Sect.~\ref{sect:implementation}.}
    \label{fig:verification}
\end{figure*}
\begin{figure}[!th]
    \centering
\includegraphics[width=\hsize]{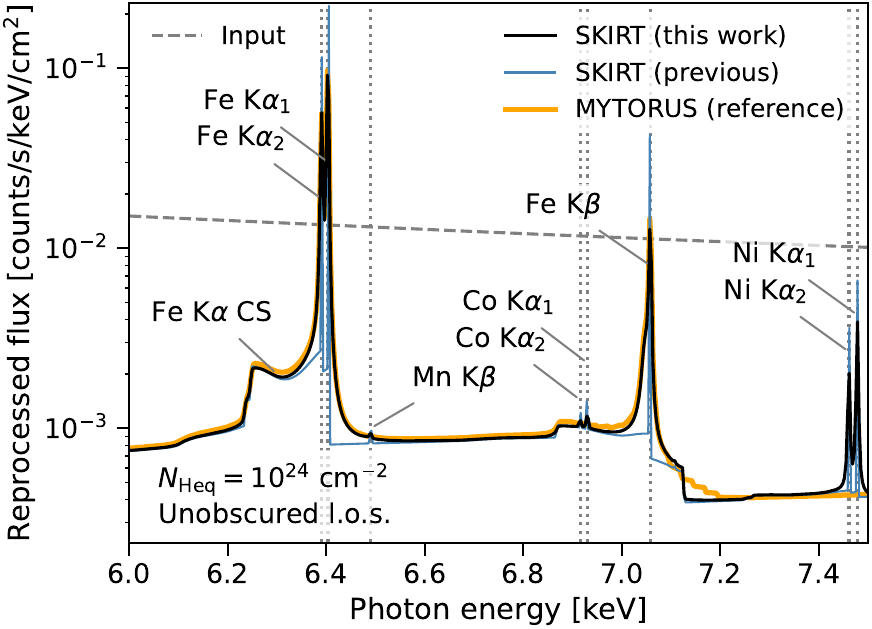}
    \label{fig:benchmark}
    \caption{Reprocessed photon flux for the benchmark model, representing an `AGN torus'. The new SKIRT implementation (black) is compared to the previous SKIRT implementation (blue) and to MYTORUS (yellow).}
\end{figure}
\section{Verification}
\label{sect:verification}
To verify the implementation of the new intrinsic line profiles, we compared the simulation results of an optically thin SKIRT model (which ensures that the intrinsic line profiles are not altered by radiative transfer effects) to the laboratory data of \citet{holzer97}, which have been corrected for instrumental effects\footnote{For Mn $\mathrm{K}\alpha$, we used the CALDB data based on \citet{eckart16}.}. We considered a SKIRT model with a ring torus geometry of cold gas, centred on an X-ray point source ($\Gamma = 1.8$), which is observed at an inclination of $90^\circ$. We adopted an equatorial column density of $N_\text{H} = 10^{22}~\text{cm}^{-2}$, which corresponds to a line-of-sight optical depth of $\tau< 4\times10^{-2}$ over the energy range of the strongest lines (i.e.\ $5.4$ to $8.3~\text{keV}$). We considered photo-absorption with \citet{anders89} abundances and no dust as well as fluorescence and bound-electron scattering, and ran the simulation with $5 \times 10^9$ photon packets.

Figure~\ref{fig:verification} shows the simulated line profiles of the six strongest fluorescent lines (in black), which are continuum-subtracted and normalised to an integrated line flux of 1. These SKIRT results are in excellent agreement with the intrinsic line profiles as measured in the laboratory (red), validating the SKIRT implementation and the data corrections discussed in Sect.~\ref{sect:implementation}. The asymmetries in the line profiles are evident from Fig.~\ref{fig:verification}.

To verify the consistency of the new implementation with the previous SKIRT implementation (which assumed infinitely narrow lines), we used a benchmark described in Sect.~4.2.1.\ of \citet{vandermeulen23}. This benchmark compares the SKIRT and MYTORUS \citep{murphy09} codes, which is particularly useful as the intrinsic line profiles of Fe K$\alpha$ and Fe K$\beta$ are also implemented in MYTORUS\footnote{\url{http://mytorus.com/mytfllfiles.html}} \citep{yaqoob24}. We reproduced the MYTORUS model in SKIRT with a uniform ring torus of cold gas, centred on an X-ray point source ($\Gamma=1.8$). Mimicking the MYTORUS setup, we assumed free-electron scattering, \citet{anders89} abundances, and no dust. We adopted an equatorial hydrogen column density of $10^{24}~\text{cm}^{-2}$ and calculated the X-ray spectrum for an unobscured sightline ($i = 45^{\circ}$). We specifically focused on the $6.0$ to $7.5~\text{keV}$ range, which contains the three most prominent fluorescent lines (Fe~K${\alpha}$, Fe~K${\beta}$, and Ni~K${\alpha}$), and the Fe~K${\alpha}$ Compton shoulder. The benchmark results are shown in Fig.~\ref{fig:benchmark}.

Apart from the distinct Fe and Ni line shapes, we find that the Fe K${\alpha}$ Compton shoulder is now slightly smoother compared to the previous SKIRT implementation, which is predictable since the Fe K${\alpha}$ line is no longer sharply defined. Furthermore, we find that the high-energy tail of the Fe~K${\beta}$ line extends to the Fe~K absorption edge at $7.124~\text{keV}$, which might complicate measurement of the absorbed continuum, especially when there is additional velocity broadening. The Fe~K${\beta}$ line asymmetry is particularly prominent.

The reprocessed continuum level away from the extended lines is consistent with MYTORUS (see Fig.~\ref{fig:benchmark}). We find a small discrepancy between $7.1$ and $7.2~\text{keV}$, which is related to the broader Fe K edge in the Compton-scattered continuum of MYTORUS. Furthermore, we find an excellent agreement between the Fe K$\alpha$ and Fe K$\beta$ line profiles as calculated with SKIRT and MYTORUS, whereas SKIRT does also predict a prominent Ni K$\alpha$ line at $\sim$$7.47~\text{keV}$. We measured the line integrals of the Fe K${\alpha}$, Fe K${\beta}$, and Ni K${\alpha}$ lines as calculated with SKIRT and find that the line fluxes agree with the previous implementation within margins of 1\%, 1\%, and 3\%, respectively. These minor differences likely arise from the continuum subtraction, which is now more ambiguous due to the extended line wings. Therefore, the new implementation is consistent with the previous SKIRT version and MYTORUS, while also incorporating intrinsic line profiles for 12 fluorescent lines, which has a significant effect on the X-ray spectrum between $6.0$ and $7.5~\text{keV}$.

\begin{figure*}
    \centering
        \includegraphics[width=\hsize]{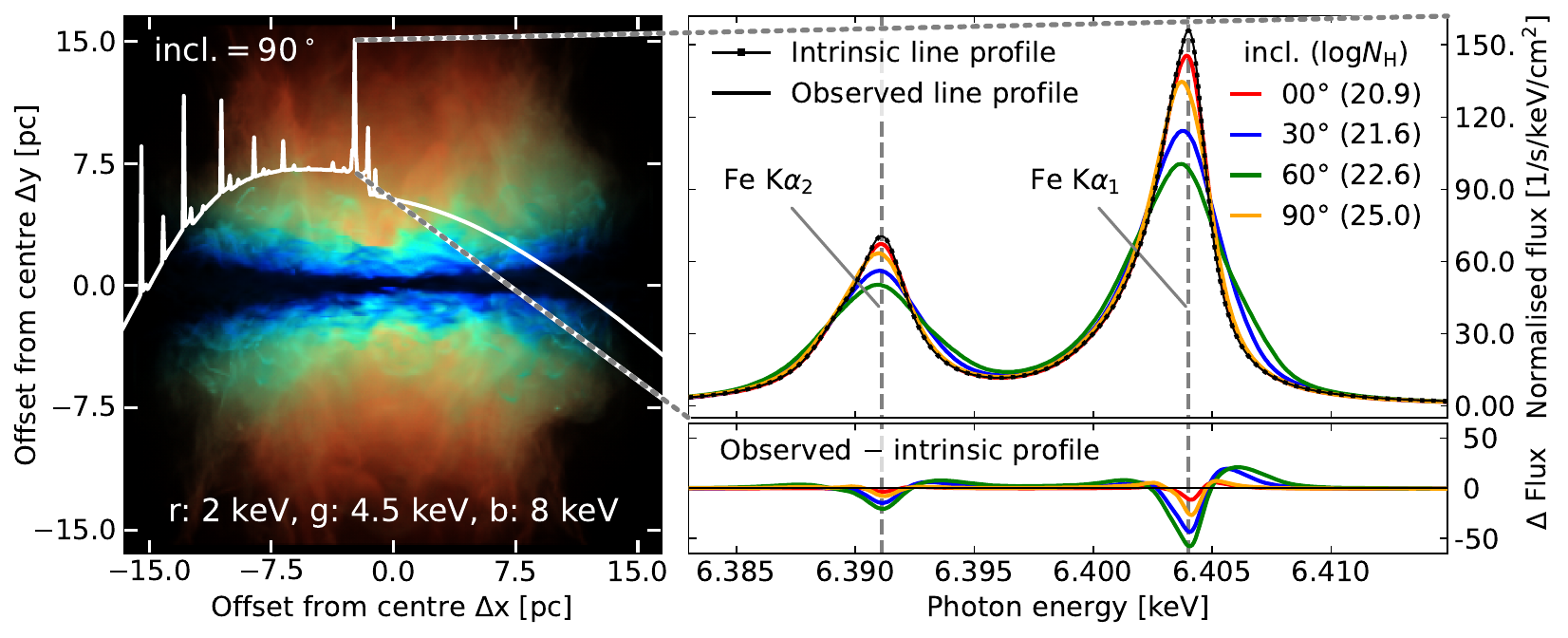}
    \caption{SKIRT radiative transfer modelling of the Fe K${\alpha}$ line in the radiation-driven fountain model of \citet{wada16}, representing the cold gas around the Circinus AGN. Right: Normalised Fe K${\alpha}$ line profiles for different inclinations. The velocity broadening of the Fe K${\alpha}$ line is significant and can only be modelled when the intrinsic line profiles are taken into account. Left: Broadband spectrum and red-green-blue image for $i=90^\circ$.}
    \label{fig:demonstration}
\end{figure*}

\section{Demonstration in 3D}
\label{sect:demonstration}
We demonstrated the new capabilities of SKIRT to predict line profiles for complex 3D models (with kinematics) by post-processing the hydrodynamical `torus' simulations of \citet{wada16}, which represent the circumnuclear gas and dust around the nearby AGN in the Circinus galaxy. This spatial distribution of cold gas, as well as the corresponding velocity field, can be directly imported into SKIRT to form a truly 3D transfer medium. 

In SKIRT, we adopted a central point source to illuminate the imported medium, with a power-law spectrum representative for the Circinus AGN ($\Gamma=1.8$ and $E_\textrm{cut}=200~\text{keV}$). We considered photo-absorption, fluorescence, and bound-electron scattering in the cold gas, with \citet{anders89} abundances and no dust. We also imported the 3D distribution of hot electrons but find that these electrons have no observable effect on the X-ray spectrum as their densities are too low ($\log N_\text{e}<22.9$ for all sightlines). In this Letter, we focus on the shape of the Fe~K${\alpha}$ line. In future work we will present more details on the X-ray post-processing of this radiation-driven fountain model with SKIRT, and an in-depth study of the X-ray observables.

The right-hand panel of Fig.~\ref{fig:demonstration} shows the normalised Fe~K${\alpha}$ line profiles as calculated with SKIRT for different inclination angles. The corresponding line-of-sight $N_\text{H}$ is given in the legend. In the left-hand panel, the broadband X-ray spectrum and a red-green-blue X-ray image at $i=90^\circ$ are shown as an illustration. The Fe~K${\alpha}$ line profiles ($i<90^\circ$) are clearly affected by velocity broadening, to an extent that should be detectable with XRISM/Resolve \citep{kilbourne24}. Yet, the observed line widths are still primarily dominated by the intrinsic line profile (dotted line), which demonstrates that these velocity measurements can only be viable when the intrinsic line profiles are taken into account.

The velocity broadening of the Fe~K${\alpha}$ lines is consistent with the Keplerian velocity of the innermost region of the equatorial torus. Indeed, by imaging the Fe~K${\alpha}$ line, we confirm that most of the line photons originate from this inner region. As a result, the line width as a function of inclination, which increases from $i=0^\circ$ to $i=60^\circ$, can be explained by the projected velocity of the equatorial gas (Fig.~\ref{fig:demonstration}). At $i=90^\circ$, the inner high-velocity component of the torus is completely obscured, resulting in a narrow line profile that is mostly produced by cold gas in the polar region (see Fig.~\ref{fig:demonstration}, left).
We note that the double-horn signature associated with the Keplerian rotation of the line-emitting gas is completely washed out by the relatively broad intrinsic line profile. 

In addition to this velocity broadening, the Fe~K${\alpha}$ line peaks are slightly redshifted (up to $0.4~\textrm{eV}$), which traces the asymmetric obscuration of the inner flow. This is further complicated by the presence of radiation-driven outflows, failed winds, and the asymmetry of the intrinsic line profile. These velocity shifts could be detected with XRISM/Resolve, which might achieve a line centroid accuracy of about $0.1~\textrm{eV}$ (at best), as evidenced by observations of the onboard ${}^{55}\mathrm{Fe}$ calibration source \citep{miller24, williams24}. The line profile for $i=90^\circ$ could be particularly interesting, as the inner obscuring disc of the Circinus AGN might be close to edge-on \citep[see e.g.][]{stalevski17, stalevski19, uematsu21, ogawa22}. In this case, we predict an Fe~K${\alpha}$ line with minimal broadening relative to the intrinsic line profile, which is mostly emitted from the polar regions and can be characterised with XRISM/Resolve.

Finally, we investigated the direct\footnote{The direct effect of the radiative transfer interactions, as opposed to the effect of the the Doppler shifts that are applied at each interaction.} effect of radiative transfer processes on the observed line profiles by re-running the SKIRT post-processing procedure without any kinematics. This time, we find that the simulated line profiles cannot be distinguished from the intrinsic line profile, which demonstrates that radiative transfer processes such as photo-absorption and bound-electron scattering do not (directly) affect the observed line shapes. This was predictable since the energy dependence of the absorption and scattering cross-sections is negligible over the energy range of a single line \citep{vandermeulen23}, while inelastic (bound-)Compton scattering produces spectral features on energy scales that are much more extended than the line profile (see for example the Fe and Ni Compton shoulders in Fig.~\ref{fig:benchmark}). However, we note that 3D radiative transfer simulations remain essential for predicting the fluorescent line strengths (together with the reprocessed continuum) and for self-consistently incorporating the 3D kinematics into the radiative transfer results.

For the same reason, the effect of dust cannot be observed in the Fe~K$\alpha$ line profile. The energy dependence of the dust extinction cross-sections is negligible over the energy range of a single line \citep{vandermeulen23}, while atoms locked in dust grains cannot contribute more than $0.35\%$ to the total bound-electron scattering \citep[i.e.\ when all Si atoms are locked in dust grains; for a dust-to-metal mass ratio of $0.3$, see Sect.~3.3.1.\ of][]{vandermeulen23}. Finally, \citet{ricci23} demonstrate that the isotropy of fluorescence is not disturbed via self-absorption in solid dust grains. We experimented with various dust fractions in SKIRT but find no observable effect on the (normalised) Fe K$\alpha$ line profile.

We find that the Fe K$\alpha$ line width is primarily caused by intrinsic broadening, with some additional broadening due to the gas kinematics. Next, we considered any further broadening that could be caused by the mixing of K$\alpha$ lines of different Fe ions, as opposed to the fully neutral Fe that is adopted in SKIRT. We estimated the charge state distribution of Fe as a function of the ionisation parameter, $\xi,$ using XSTAR \citep{kallman04} and calculated the total observed Fe K$\alpha$ line centroid and its spread based on the ion line energies quoted in \citet{yamaguchi14}. Up to $\xi=10\text{~erg~cm~s}^{-1}$, the line broadening due to the mixing of quasi-neutral Fe K$\alpha$ lines is less than $5~\text{eV}$, while it would be less than $0.5~\text{eV}$ for $\xi<1\text{~erg~cm~s}^{-1}$. For the 3D torus model by \citet{wada16}, the radiation-driven gas has an ionisation parameter $\xi<1\text{~erg~cm~s}^{-1}$, whereas $\xi<10^{-4}\text{~erg~cm~s}^{-1}$ for the dense equatorial disc producing most of the Fe K$\alpha$ line photons \citep{ogawa22}. Therefore, this broadening would be less than $0.5~\text{eV}$ (i.e.\ below the $5~\text{eV}$ resolution of XRISM/Resolve) and can be safely neglected.

\section{Summary and outlook}
\label{sect:summary}
This Letter reports on the implementation of the intrinsic line profiles of 12 fluorescent lines
into the 3D radiative transfer code SKIRT (Sect.~\ref{sect:implementation}) and presents the Fe K${\alpha}$ line profiles for a 3D torus model that also includes kinematics (Sect.~\ref{sect:demonstration}).

We find that the SKIRT results in the optically thin regime match the intrinsic line profiles as measured in the laboratory (Fig.~\ref{fig:verification}), and we confirm the consistency of the new results with the previous SKIRT implementation, away from these lines (Fig.~\ref{fig:benchmark}). This comparison further highlights the importance of the new line profiles, with prominent Fe K${\alpha}$, Fe K${\beta}$, and Ni K${\alpha}$ lines showing their extended, asymmetric line wings and a smoothed Fe K${\alpha}$ Compton shoulder (Fig.~\ref{fig:benchmark}).

In Sect.~\ref{sect:demonstration} we study the formation of Fe K${\alpha}$ line profiles in the 3D AGN torus model of \citet{wada16}, which also includes kinematics. We find that the intrinsic line profiles are broadened and shifted to an extent that should be detectable with XRISM/Resolve (Fig.~\ref{fig:demonstration}), while also demonstrating that the observed line profiles are still dominated by the intrinsic Fe K${\alpha}$ profile. We conclude that velocity measurements on fluorescent lines are only viable when the intrinsic line profiles are taken into account. Finally, we find that the fluorescent line profiles directly trace the cold-gas kinematics without being affected by photo-absorption or (bound-)Compton scattering.

To our knowledge, there is no other publicly available X-ray radiative transfer framework with sufficient flexibility that can model the effect of 3D velocities during the radiative transfer routine such that gas kinematics are self-consistently incorporated into the simulation results (as opposed to some smoothing in post-processing). Therefore, the SKIRT code could constitute a unique tool for studying the kinematics of cold gas through X-ray spectroscopy now that the intrinsic line profiles of the strongest fluorescent lines are properly incorporated into the code.

With the advent of the first XRISM/Resolve data, this update to the X-ray framework of SKIRT is timely and will be crucial for constraining the 3D density and velocity structure of cold gas from X-ray microcalorimeter spectra. The SKIRT code is publicly available and well documented\footnote{\url{https://skirt.ugent.be}}, and the community is warmly invited to use the code in any way they see fit.

\begin{acknowledgements}
      We wish to thank the anonymous referee for their careful reading and constructive comments. B.\ V.\ acknowledges support by the Fund for Scientific Research Flanders (FWO-Vlaanderen, project 11H2121N). Numerical computations were carried out on Cray XC50 at Center for Computational Astrophysics, National Astronomical Observatory of Japan. This work is supported by the Japan Society for the Promotion of Science (JSPS) KAKENHI grant No. JP21H04496 (K.\ W.).
\end{acknowledgements}
\bibliographystyle{aa}
\bibliography{finalrefs}

\end{document}